\documentclass[prb,twocolumn]{revtex4}
\usepackage{graphicx}
\begin{document}
\title{Quantum pumping and rectification effects in Aharonov-Bohm-Casher ring-dot systems}
\author{F. Romeo, R. Citro and M. Marinaro}
\affiliation{Dipartimento di Fisica ``E. R. Caianiello'',
C.N.I.S.M. and NANOMATES  (Research Centre for NANOMAterials and
nanoTEchnology), Universit{\`a} degli Studi di Salerno, Via S.
Allende, I-84081 Baronissi (Sa), Italy}

\begin{abstract}
We study the time-dependent transport of charge and spin through a
ring-shaped region sequentially coupled to a weakly interacting
quantum dot in the presence of an Aharonov-Bohm flux and
spin-orbit interaction. The time-dependent modulation of the
spin-orbit interaction, or of the corresponding Aharonov-Casher
flux, together with the modulation of the dot-level induces an
electrically pumped spin current even in absence of a charge
current. The results beyond the adiabatic regime show that an
additional rectification current proportional to $\cos(\varphi)$,
being $\varphi$ the relative phase between the time varying
parameters, is generated. We discuss the relevance of such term in
connection with recent experiments on out-of-equilibrium quantum
dots.
\end{abstract}


\maketitle
\section{Introduction}
In recent years the process of miniaturization of manmade
electronic circuits has permitted to reach the molecular scale
providing the opportunity of testing quantum mechanics in
nano-electronics measurements\cite{review}. In this framework, one
of the most exciting challenges is to encode information by means
of the electron spin instead of the charge, giving rise to the
so-called spin-based electronics or spintronics\cite{loss,sarma}.
To face the need of spin-polarizing systems acting as a source in
spintronics devices, one promising possibility is to exploit the
quantum interference effects by external electric or magnetic
fields. In ring-shaped structures made of semiconducting materials
a spin sensitive phase, the Aharonov-Casher
phase\cite{Aharonov-Casher84},  is originated by the Rashba
spin-orbit interaction\cite{rashba60, bychkov-rashba84}. Such
phase combined with the Aharonov-Bohm phase\cite{Aharonov-Bohm59}
induced by a magnetic field is an useful tool to achieve
spin-polarizing devices\cite{governale,frustaglia_ring}. Another
interesting phase interference effect is originated by the
periodic modulation of two out-of-phase parameters affecting the
scattering properties of a nanostructure. Such phase effect, known
in the literature as quantum pumping, was first introduced by
Thouless\cite{thouless83}. After the Thouless theory, a scattering
approach to the adiabatic quantum pumping was formulated by P. W.
Brouwer \cite{brouwer98} who showed that the d.c. current pumped
by means of an adiabatic modulation of two out-of-phase
independent parameters can be expressed in terms of the parametric
derivatives of the scattering matrix. In the adiabatic regime
described by the Brouwer formula, i.e. when the pumping frequency
is much slower than the tunneling rates, a d.c. current
proportional to $\omega \sin(\varphi)$ is originated, being
$\varphi$ the phase difference between the two parameters. Such
theoretical prediction has been verified experimentally by M.
Switkes et al.\cite{switkes99} even though some anomalies in the
current-phase relation have been reported. In particular, it has
been observed a non-vanishing current at $\varphi=0$. Several
anomalies observed in the experiment can be explained by
rectification of a.c. displacement currents as proposed in
Ref.[\onlinecite{brouwer-rect-01}]. According to this work, the
rectified currents are responsible for measurable effects which
may be dominant over the pumping currents. In order to
discriminate between rectified currents and pumping effects
symmetry arguments can be exploited. For instance, the d.c.
rectification voltage $V_{rect}$ is symmetric under reversal of
the magnetic field $V_{rect}(B)=V_{rect}(-B)$, while the voltage
generated by a quantum pump is not. On the other hand, it has been
shown in Ref.[\onlinecite{wang-wang-guo02}] that finite frequency
effects, considered within a non-equilibrium Green's function
approach, can lead to current-phase relations of the form $I_c
\sim a \sin(\varphi)+b \cos(\varphi)$, where the coefficients $a$
and $b$ are function of the pumping frequency $\omega$.
Differently from the adiabatic regime where the pumped currents
are odd function of the relative phase $\varphi$ between the
modulated parameters, any symmetry can be realized in the general
non-equilibrium case.

\begin{figure}[htbp]
\centering
\includegraphics[scale=.45]{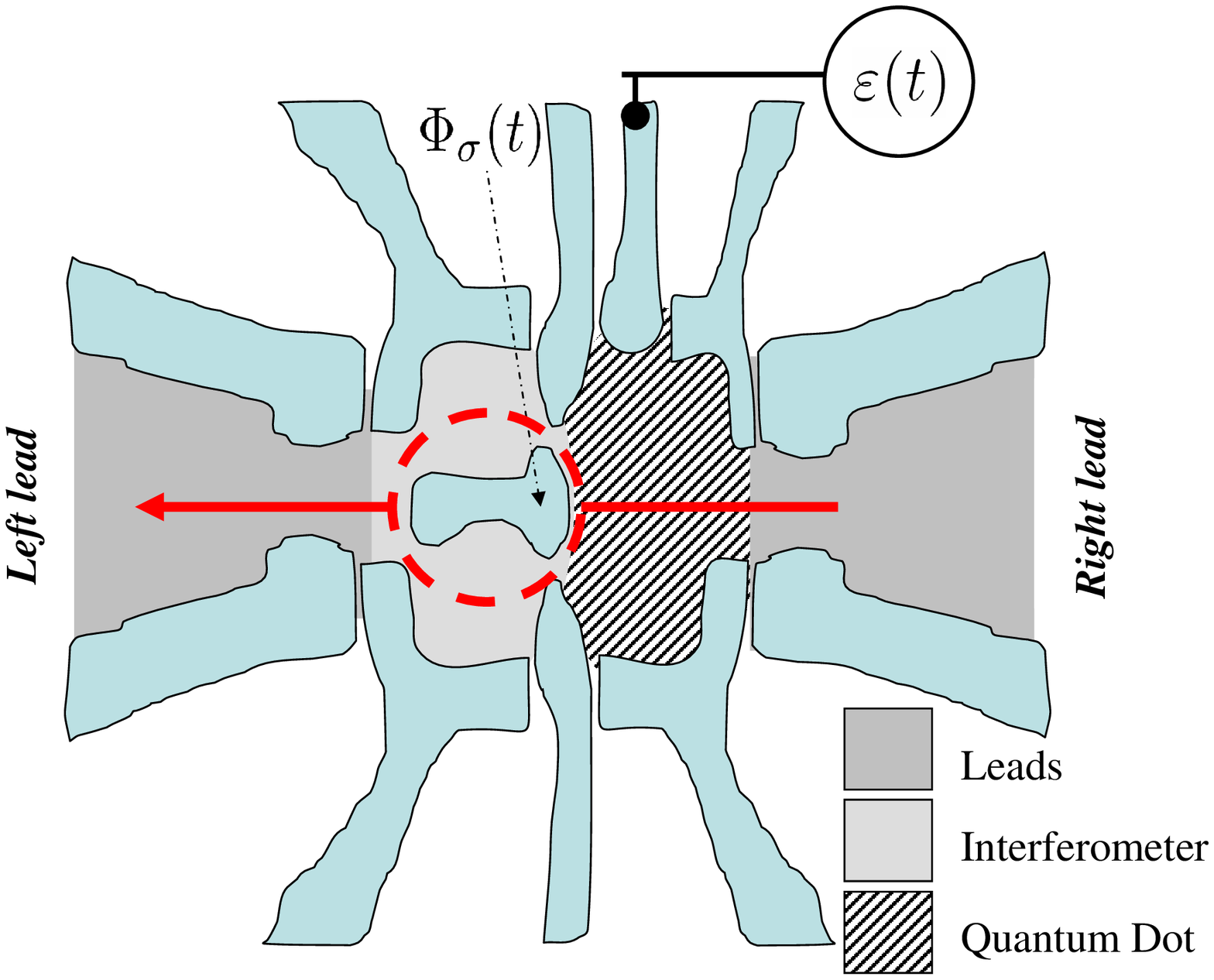}
\caption{An Aharonov-Bohm-Casher quantum ring sequentially coupled
to a quantum dot. The energy on the dot and the Aharonov-Casher
flux are modulated in time with frequency $\omega$.}
\label{fig:device}
\end{figure}

In the following we analyze the charge and spin pumping in a
ring-shaped conductor sequentially coupled to a quantum dot (see
Fig.\ref{fig:device}) and apply the non-equilibrium Green's
function approach to analyze the dc current from the adiabatic to
non-adiabatic regime addressing the question about the existence
of rectification terms. In the ring region shown in
Fig.\ref{fig:device} the electrons feel an Aharonov-Bohm phase
associated to a time varying Aharonov-Casher phase. The last is
related to the Rashba spin-orbit interaction which is tunable by
means of a gate voltage \cite{nitta-97}. An additional
time-dependent modulation of the dot energy level is also
considered . If no voltage bias is present between the two
external leads, the electron current is activated by absorption
and emission of quantized photon energy. Thus, in the following
the charge and spin pumped current are studied as a side-effect of
boson-assisted
tunneling. \\
The paper is organized as follows. In section \ref{sec:model} we
introduce the  model Hamiltonian and derive the general expression
for the non-equilibrium Green's function and respective
self-energies for the non-interacting and weakly interacting case.
In section \ref{sec:one-ph-approx-curr}, we employ a
one-photon-approximation  and obtain a compact expression for the
d.c. current pumped in the left lead.  In section
\ref{sec:results}, we present the results of our analysis as a
function of the phase and interaction effects. Finally, in section
\ref{sec:conclusions} some conclusions are given.
\section{The model and non-equilibrium current}\label{sec:model}
The Hamiltonian of an Aharonov-Bohm-Casher ring sequentially
coupled to an interacting quantum dot in the presence of time
varying parameters can be written, in the local spin frame, as
follows\cite{nota-hamiltoniana}:
\begin{eqnarray}\label{eq:hamiltonian1}
H(t)&=& H_c+\sum_{\sigma} \varepsilon(t) d^{\dag}_{\sigma}
d_{\sigma}+U n_{\uparrow}n_{\downarrow}+\nonumber \\ &+&
\sum_{k\sigma}[2u \cos(\Phi_{\sigma}(t)) d^{\dag}_{\sigma}c_{k
\sigma l }+w c^{\dag}_{k \sigma r}d_{\sigma}]+H.c.,
\end{eqnarray}
where $H_c=\sum_{k \sigma, \alpha=\{l,r\}}\varepsilon^\alpha_k
c^{\dag}_{k \sigma \alpha }c_{k \sigma \alpha}$ is the free
electrons Hamiltonian describing the left/right ($l/r$) leads kept
at the same chemical potential $\mu$. The second and third term
represent the dot Hamiltonian consisting of the electron-electron
interaction term $U n_{\uparrow}n_{\downarrow}$
($n_{\sigma}=d^{\dag}_{\sigma}d_{\sigma}$) and of the time
dependent dot energy level
$\varepsilon(t)=\varepsilon_0+\varepsilon_\omega \sin(\omega
t+\varphi)$, being $\omega$ the frequency of the modulation. The
last term in Hamiltonian describes the tunneling between the left
lead and the dot, $u \cos(\Phi_{\sigma})$, and the right lead and
the dot through a tunnel barrier, $w$. The transmission
coefficients $u$ and $w$, which in general may be spin and
momentum dependent, are considered here constant for simplicity,
i.e. $u\approx u(k=k_F)$ and $w\approx w(k=k_F)$, $k_F$ being the
Fermi momentum.\\
The electrons coming from the left lead acquire the time-dependent
spin sensitive phase $\Phi_{\sigma}(t)=\pi (\Phi_{AB}+\sigma
\Phi_R(t))$ ($\sigma=\pm 1$), where $\Phi_{AB}$ is the
Aharonov-Bohm phase, while  $\Phi_{R}$ is the Aharonov-Casher
phase produced by the modulation of the Rashba spin-orbit
interaction on the ring
$\Phi_R(t)=\Phi_R^0+\Phi_R^{\omega}\sin(\omega
t)$[\onlinecite{frustaglia_ring}]. The Hamiltonian given in
Eq.\ref{eq:hamiltonian1} can be rewritten by means of a plane wave
expansion in the following form:
\begin{eqnarray}
&&H(t)= H_0+\varepsilon_\omega \sin(\omega
t+\varphi)\sum_{\sigma}d^{\dag}_{\sigma}d_{\sigma}+\nonumber \\
&&+\sum_{k\sigma}4u[\cos(\Phi^0_{\sigma})A(t)-\sigma
\sin(\Phi^0_{\sigma})B(t)] d^{\dag}_{\sigma}c_{k\sigma l}+H.c.,
\end{eqnarray}
where the static part $H_0$ of the Hamiltonian is the same as in
Eq.(\ref{eq:hamiltonian1}) with: $\varepsilon(t)\rightarrow
\varepsilon_0$, $u \rightarrow u J_0(\pi \Phi_R^\omega)$,
$\Phi_R(t)\rightarrow \Phi_R^0$, while the functions $A(t)$ and
$B(t)$ are given by (see APPENDIX \ref{app:bessel-expansion}):
\begin{eqnarray}
A(t)&=&\sum_{n=1}^{\infty}J_{2n}(\pi\Phi_R^\omega)\cos(2n \omega
t)\\
B(t)&=&\sum_{n=1}^{\infty}J_{2n-1}(\pi
\Phi_R^\omega)\sin((2n-1)\omega t),
\end{eqnarray}
where $J_n(x)$ are the Bessel functions of first kind. In absence
of an external voltage, the quantum transport of particles through
the structure is only due to absorption/emission of energy quanta
$\hbar \omega$ associated to the external time-dependent fields.

To calculate the pumped current we employ the Keldysh Green's
functions technique as formulated in
Ref.[\onlinecite{jauho-wingreen-meir94}]. The current in the left
lead is given by $I_l(t)=-e\frac{d}{dt}\langle
N_l\rangle=-\frac{ie}{\hbar}\langle[H,N_l]\rangle$
($N_l=\sum_{k\sigma}c^{\dag}_{k\sigma l}c_{k\sigma l}$). It can be
rewritten in terms of the retarded and lesser Green's functions
$G^{r/<}_{\sigma\sigma'}(t,t_1)$ and of the advanced/retarded
self-energies $\Sigma^{a/<}_{l,\sigma\sigma'}(t_1,t)$ of the
quantum dot according to the following expression:
\begin{eqnarray}\label{eq:current}
I_l(t)&=&\sum_{\sigma}I_l^{\sigma}(t)\\
I_l^{\sigma}(t)&=& \frac{2e}{\hbar}\sum_{\sigma'} Re \{ \int dt_1
G^{r}_{\sigma\sigma'}(t,t_1)\Sigma^{<}_{l,\sigma'\sigma}(t_1,t)+ \nonumber \\
&+&
G^{<}_{\sigma\sigma'}(t,t_1)\Sigma^{a}_{l,\sigma'\sigma}(t_1,t)\}.
\end{eqnarray}
For a time-dependent problem the retarded, advanced end lesser
($r$,$a$,$<$) Green's functions and respective self-energies
depend explicitly on two time variables instead of one. Thus,
employing the two time Fourier transform (see APPENDIX
\ref{app:fourier}) the current (\ref{eq:current}) can be rewritten
as:
\begin{eqnarray}\label{eq:current-fourier}
I_l^{\sigma}(t)&=& \frac{2e}{\hbar}\sum_{\sigma'}Re\{ \int
\frac{dE_1 d E_2 dE_3}{(2\pi)^3}e^{i(E_3-E_1)t}\times \nonumber \\
&&\times[G^{r}_{\sigma\sigma'}(E_1,E_2)\Sigma^{<}_{l,\sigma'\sigma}(E_2,E_3)+
\nonumber \\
&&+G^{<}_{\sigma\sigma'}(E_1,E_2)\Sigma^{a}_{l,\sigma'\sigma}(E_2,E_3)]\},
\end{eqnarray}
where the lesser Green's function $\textbf{G}^{<}(E_1,E_2)$ of the
dot is given by the Keldysh equation:
\begin{equation}
\label{eq:keldysh}
\mathbf{G}^{<}(E_1,E_2)= \int\frac{d\xi_1
d\xi_2}{(2\pi)^2}\mathbf{G}^{r}(E_1,\xi_1)
\mathbf{\Sigma}^{<}(\xi_1,\xi_2)\mathbf{G}^{a}(\xi_2,E_2),
\end{equation}
and the following relation between the retarded and advanced
quantities can be used:
$\Xi^{a}(E_1,E_2)=[\Xi^{r}(E_2,E_1)]^{\dag}$, where $\Xi=G$ or
$\Sigma$. In order to compute the current the knowledge of $G^r$,
$\Sigma^r$, $\Sigma^<$ is required. Below we are going to
calculate them for the non-interacting dot and the weakly
interacting one.
In both cases, the wide band limit (WBL) will be employed for
simplicity.

\subsection{Non-interacting case ($U=0$)}
In the $U=0$ case the expression of the self-energies can be
obtained exactly. By calling $w=\gamma u$ ($\gamma \in
\mathbf{R}$),  the retarded and lesser self-energies can be
expressed in terms of the corresponding Green's functions of the
leads, namely
$[g^{r}_{k\alpha}(t,t')]_{sp}=-i\delta_{sp}\theta(t-t')e^{-i\varepsilon^{\alpha}_k(t-t')}$
and
$[g^{<}_{k\alpha}(t,t')]_{sp}=i\delta_{sp}f(\varepsilon^{\alpha}_k)e^{-i\varepsilon^{\alpha}_k(t-t')}$,
by the following relations:
\begin{eqnarray}
&&\Sigma^{r,<}_{sp}(t_1,t_2)=\sum_{k,\alpha \in r}\gamma^2|u|^2
[g^{r,<}_{kr}(t_1,t_2)]_{sp}+ \nonumber \\ &+& \sum_{k,\alpha \in
l}4 |u|^2
\cos(\Phi_s(t_1))\cos(\Phi_p(t_2))[g^{r,<}_{kl}(t_1,t_2)]_{sp},
\end{eqnarray}
where $s,p$ are spin indices ($\uparrow,\downarrow$),
$f(\varepsilon_k)$ is the Fermi function, while
$\varepsilon^{l}_k=\varepsilon^{r}_k=\varepsilon_k$ in absence of
voltage bias. In WBL limit and making the substitution
$\sum_{k,\alpha}\rightarrow \int d\varepsilon
\rho_\alpha(\varepsilon)$, we get:
\begin{eqnarray}
\Sigma^{r}_{sp}(t_1,t_2)&=&-i\delta_{sp}\delta(t_1-t_2)\Gamma^0
\lbrack\gamma^2/2+2\cos^2(\Phi_s(t_1)) \rbrack \nonumber \\
\Sigma^{<}_{sp}(t_1,t_2)&\approx&
i\delta_{sp}f(t_1-t_2)\Gamma^0\lbrack\gamma^2+4\cos^2(\Phi_s(t_1))\rbrack,
\end{eqnarray}
where we introduced the quantities $\Gamma^0=2\pi \rho |u|^2$ and
$f(t_1-t_2)=\int
\frac{d\varepsilon}{2\pi}f(\varepsilon)e^{-i\varepsilon(t_1-t_2)}$.
By defining the left and right transition rates
$\Gamma_s^l(t_1)/2=2\Gamma^0\cos^2(\Phi_s(t_1))$ and
$\Gamma_s^r(t_1)/2=\Gamma^0\gamma^2/2$, the retarded self-energy
can be written as
$\Sigma^{r}_{sp}(t_1,t_2)=-i\delta_{sp}\delta(t_1-t_2)\{\Gamma_s^l(t_1)/2+\Gamma_s^r(t_1)/2\}$
and thus the retarded Green's function of the quantum dot takes
the following form \cite{langreth91}:
\begin{eqnarray}
G^r_{sp}(t,t')&=&g^r_{sp}(t,t')\times \nonumber \\
&\times&\exp\Bigl\{-\frac{1}{2}\int^t_{t'}dt_1[\Gamma^{l}_s(t_1)+\Gamma^{r}_s(t_1)]\Bigl\},\nonumber \\
g_{sp}^r(t,t')&=&-i\delta_{sp}\theta(t-t')\exp\Bigl\{-i\int^{t}_{t'}
dt_1 \varepsilon(t_1) \Bigl\}.
\end{eqnarray}
It can be computed exactly after having performed the following
plane-wave expansion of the retarded self-energy:
\begin{eqnarray}
\label{eq:self_energy_ret}
\Sigma^{r}_{sp}(t_1,t_2)&=&-i\delta_{sp}\delta(t_1-t_2)\Gamma^0\lbrack
1+\gamma^2/2+\nonumber \\
&+&J_0(2\pi\Phi^{\omega}_R)\cos(2\Phi^0_s)+\nonumber \\
&+&\cos(2\Phi^0_s)\mathcal{M}(t_1)-s
\sin(2\Phi^0_s)\mathcal{N}(t_1)\rbrack,
\end{eqnarray}
where the following auxiliary functions have been introduced:
\begin{eqnarray}
\mathcal{M}(t)&=&2\sum_{n=1}^{\infty}J_{2n}(2\pi\Phi_R^\omega)\cos(2n
\omega
t)\\
\mathcal{N}(t)&=&2\sum_{n=1}^{\infty}J_{2n-1}(2\pi
\Phi_R^\omega)\sin((2n-1)\omega t),
\end{eqnarray}
and whose Fourier transform is given by:
\begin{eqnarray}
\label{eq:sigma_r_ft}
&&\Sigma^{r}_{sp}(E_1,E_2)=-\pi\delta_{sp}\{i\delta(E_1-E_2)\mathcal{Q}_1^{s}+
\nonumber \\&&+i\mathcal{Q}_2^{s}\sum_{\eta=\pm
1,n=1}^{\infty}J_{2n}(2\pi\Phi_R^{\omega})\delta(E_1-E_2+2n\omega\eta)-\mathcal{Q}_3^{s}\times
\nonumber \\&&\times\sum_{\eta=\pm 1, n=1}^{\infty}\eta
J_{2n-1}(2\pi\Phi_R^{\omega})\delta(E_1-E_2+(2n-1)\omega\eta)\}.
\nonumber \\
\end{eqnarray}
An analogous plane-wave expansion can be performed for the lesser
self-energy $\Sigma^{<}_{sp}(t_1,t_2)$, whose Fourier transform is
the following:
\begin{eqnarray}
\label{eq:self_energy_less} &&\Sigma^{<}_{sp}(E_1,E_2)=2\pi i
\delta_{sp}\{ f(E_1)\delta(E_1-E_2)\mathcal{Q}^{s}_1+\nonumber \\
&&+\sum^{\infty}_{\eta=\pm 1, n=1}\mathcal{Q}^{s}_2
J_{2n}(2\pi\Phi^{\omega}_R)\times \nonumber \\ &&\times
f(E_1+2n\omega\eta) \delta(E_1-E_2+2n\omega \eta)+ \nonumber \\
&&+\sum^{\infty}_{\eta=\pm 1, n=1}i\mathcal{Q}^{s}_3\eta
J_{2n-1}(2\pi\Phi^{\omega}_R)\times \nonumber \\ &&\times
f(E_1+(2n-1)\omega\eta)\delta(E_1-E_2+(2n-1)\omega \eta)\},
\end{eqnarray}
where $\mathcal{Q}^{s}_j$ are given by:
\begin{eqnarray}
&&\mathcal{Q}^{s}_1=2\Gamma^0\{\gamma^2/2+1+J_0(2\pi\Phi^{\omega}_R)\cos(2\Phi^0_s)\}
\nonumber \\
&&\mathcal{Q}^{s}_2=2\Gamma^0\cos(2\Phi^0_s)\nonumber \\
&&\mathcal{Q}^{s}_3=2\Gamma^0 s \sin(2\Phi^0_s).
\end{eqnarray}
The substitution of Eq.(\ref{eq:self_energy_less}) and of
$\textbf{G}^{r}(E_1,E_2)$   in Eq.(\ref{eq:keldysh}) permits to
determine the $\textbf{G}^{<}(E_1,E_2)$.

The knowledge of the retarded and lesser Green's function enables
us to calculate the current generated by the pumping procedure in
the form of a trigonometric series, i.e.
$I^{\sigma}_l(t)=I_0^{\sigma}+\sum^{\infty}_{n=1}[c_n^{\sigma}\cos(n\omega
t )+s_n^{\sigma}\sin(n\omega t)]$, allowing to recognize the d.c.
component of the current.

\subsection{Weakly interacting case ($U\approx 0$)}
The weakly interacting limit ($U\approx 0$) can be studied by
means of a self-consistent Hartree-Fock theory which is known to
give suitable results when the Coulomb interaction $U$ is small
\cite{flensberg-book04,doniach-book04} (i.e., $U \ll \Gamma^0$).
In this framework, the energy of the electrons on the dot is
modified by a spin dependent term related to the occupation number
of the electron of opposite spin and thus the spin dependent
energy becomes $\varepsilon_{\sigma}(t)=\varepsilon(t)+U\langle
n_{\bar{\sigma}}(t)\rangle$ ($\bar{\sigma}=-\sigma$). The
occupation number $\langle n_{\sigma}(t)\rangle$ is calculated
self-consistently by means of the relation $i\langle
n_{\sigma}(t)\rangle=G_{\sigma\sigma}^{<}(t,t)$. The retarded
Green's function of the dot is modified by the interaction
according to the expression:
\begin{equation}
\mathcal{G}^r_{sp}(t,t')=G^r_{sp}(t,t')\exp \Bigl\{-i U
\int_{t'}^{t}dt_1\langle n_{\bar{s}}(t_1) \rangle\Bigl\},
\end{equation}
where $G^r_{sp}(t,t')$ represents the retarded Green's function
derived in the non-interacting case. In order to determine the
interacting Green's function, we can write $\langle
n_{\bar{s}}(t_1) \rangle$ as a trigonometric series of $\sin( n
\omega t)$, $\cos(n \omega t)$ with unknown coefficients
calculated in a self-consistent way, as explained  below.

\section{The single photon approximation}\label{sec:one-ph-approx-curr}
Hereon we focus on the weak pumping limit and consider only single
photon processes, i.e. involving emission or absorption of a
single energy quantum $\hbar \omega$. The weak pumping limit (i.e.
the case in which a pure $\sin(\varphi)$ behavior of the current
is expected) is very important in experiments where the higher
harmonics contribution seem to be negligible even though others
anomalies occur. Such anomalies in the current-phase relation will
be discussed here later.

\subsection{Non-interacting case ($U=0$)}
Within the single photon approximation the self energies
(\ref{eq:sigma_r_ft})-(\ref{eq:self_energy_less}) can be
approximated as:
\begin{eqnarray}
\label{eq:sigma_approx} \Sigma^{r}_{sp}(E_1,E_2)&\approx&
-\pi\delta_{sp}\{ i \delta(E_1-E_2)\mathcal{Q}^{s}_1 +\nonumber \\
&+&\mathcal{Q}^{s}_3 J_1(2\pi\Phi_{R}^{\omega})\sum_{\eta=\pm 1
}\eta\delta(E_1-E_2-\eta\omega)\} \nonumber \\
\Sigma^{<}_{sp}(E_1,E_2)&\approx&2\pi\delta_{sp}\{i
f(E_1)\mathcal{Q}_1^{s}-\mathcal{Q}^{s}_3J_1(2\pi\Phi_{R}^{\omega})\times
\nonumber \\ &\times&\sum_{\eta=\pm 1 }\eta f(E_1+\eta
\omega)\delta(E_1-E_2+\eta \omega)\}.
\end{eqnarray}
The approximation is valid for small values of $2\pi\Phi_R^\omega$
so that the inequality $J_1(2\pi\Phi_R^\omega)\gg
J_n(2\pi\Phi_R^\omega)$, $n>1$, is verified. Using the above
approximated form of (\ref{eq:sigma_approx})
the retarded Green's function of the quantum dot can be rewritten
as follows:
\begin{eqnarray}
G^{r(1)}_{sp}(t,t')&=&-i\delta_{sp}\theta(t-t')\times \nonumber \\
&\times&\exp\Bigl\{-i[\varepsilon_0-i\mathcal{Q}^{s}_1/2)](t-t')\Bigl\}\times
\nonumber \\&\times&\exp\Bigl\{-i[\Lambda_1^s
\int_{t'}^{t}dt_1\sin(\omega t_1)+\nonumber \\ &+&\Lambda_2
\int_{t'}^{t}dt_1\cos(\omega t_1)] \Bigl\},
\end{eqnarray}
where we introduced the coefficients:
\begin{eqnarray}
\Lambda_1^s&=&
\varepsilon_{\omega}\cos(\varphi)+iJ_1(2\pi\Phi_R^{\omega})\mathcal{Q}_3^s
\nonumber \\ \Lambda_2&=&\varepsilon_{\omega}\sin(\varphi),
\end{eqnarray}
and the upper index (1) stands for the single-photon
approximation. Making a further expansion of the retarded Green's
function for small $\varepsilon_{\omega}/\omega$ leads to the
result:
\begin{eqnarray}
\label{eq:gr_app_sf} G^{r(1)}_{sp}(t,t')&\simeq
&-i\delta_{sp}\theta(t-t')\times \nonumber \\
&\times&\exp\Bigl\{-i[\varepsilon_0-i\mathcal{Q}^{s}_1/2)](t-t')\Bigl\}\times
\nonumber \\&\times&\Bigl\{\Delta^s_0+\Delta^s_1
\mathcal{C}(t,t')+\Delta^s_2\mathcal{S}(t,t') \Bigl\},
\end{eqnarray}
where $\mathcal{C}(t,t')=\cos(\omega t')-\cos(\omega t)$,
$\mathcal{S}(t,t')=\sin(\omega t')-\sin(\omega t)$, while the
coefficients $\Delta^s_j$ have been defined as follows:
\begin{eqnarray}
\Delta^s_0&=&J_0\Bigl(\frac{\varepsilon_\omega\sin(\varphi)}{\omega}\Bigl)^2J_0\Bigl(\frac{\varepsilon_\omega\cos(\varphi)}{\omega}\Bigl)^2\times
\nonumber \\ &\times&
I_0\Bigl(\frac{J_1(2\pi\Phi_R^\omega)\mathcal{Q}_3^s}{\omega}\Bigl)^2
\nonumber \\
\Delta^s_1&=&2J_0\Bigl(\frac{\Lambda_2}{\omega}\Bigl)^2I_0\Bigl(\frac{Im\{\Lambda^s_1\}}{\omega}\Bigl)J_0\Bigl(\frac{Re\{\Lambda^s_1\}}{\omega}\Bigl)\times
\nonumber \\
&\times&\Bigl[J_0\Bigl(\frac{Re\{\Lambda^s_1\}}{\omega}\Bigl)I_1\Bigl(\frac{Im\{\Lambda^s_1\}}{\omega}\Bigl)-iJ_1\Bigl(\frac{Re\{\Lambda^s_1\}}{\omega}\Bigl)\times
\nonumber \\
&\times&I_0\Bigl(\frac{Im\{\Lambda^s_1\}}{\omega}\Bigl)\Bigl]
\nonumber \\ \Delta^s_2&=&2i
I_0\Bigl(\frac{Im\{\Lambda^s_1\}}{\omega}
\Bigl)^2J_0\Bigl(\frac{Re\{\Lambda^s_1\}}{\omega}\Bigl)^2J_1\Bigl(\frac{\Lambda_2}{\omega}\Bigl)J_0\Bigl(\frac{\Lambda_2}{\omega}\Bigl),
\nonumber \\
\end{eqnarray}
where $I_n(x)$ ($n=0,1$) represents the modified Bessel function
of first kind and order $n$. The above result can be conveniently
rewritten in terms of the two-time Fourier transform and thus we
have:
\begin{eqnarray}
G^{r(1)}_{sp}(E_1,E_2)&=&2\pi\delta_{sp}\Bigl\{\frac{\Delta_0^s\delta(E_1-E_2)}{\mathcal{D}^s(E_1)}+
\nonumber \\ &+&\sum_{\eta=\pm 1 }\frac{\eta \omega
\mathcal{R}^s_{\eta}\delta(E_1-E_2+\eta\omega)
}{\mathcal{D}^s(E_1)(\mathcal{D}^s(E_1)+\eta\omega)}\Bigl\},
\end{eqnarray}
where we defined
$\mathcal{R}^s_{\eta}=(\Delta_1^s-i\eta\Delta_2^s)/2$ and
$\mathcal{D}^s(E_1)=E_1-\varepsilon_0+i \mathcal{Q}^s_1/2$. The
knowledge of the retarded Green's function allows us to write the
lesser Green's function by means of the Keldysh equation in this
way:
\begin{eqnarray}
G^{<(1)}_{ss}(E_1,E_2)&=&2\pi i
\mathcal{Q}^s_1\mathcal{F}_s^{0}(E_1,E_2)+\nonumber \\ &+&2\pi
\mathcal{Q}^s_3J_1(2\pi\Phi_R^{\omega})\sum_{\eta=\pm 1
}\eta\mathcal{F}_s^{-\eta}(E_1,E_2), \nonumber \\
\end{eqnarray}
where we defined the following integral function:
\begin{eqnarray}
\mathcal{F}^{\eta}_{\sigma}(E_1,E_2)&=&\sum_{s}\int
\frac{d\xi}{(2\pi)^2}G^{r(1)}_{\sigma s}(E_1,\xi-\eta
\omega)\times \nonumber \\
&\times& f(\xi)G^{r(1)\ast}_{\sigma s}(E_2,\xi).
\end{eqnarray}
The above function, disregarding terms quadratic in $J_1(x)$ and
$I_1(x)$ and additional terms describing higher order processes
(roughly cubic in $\sim [\mathcal{D}^s(E_1)]^{-1}$), can be
written in the simple form:
\begin{eqnarray}
\mathcal{F}^{\eta}_{s}(E_1,E_2)=\frac{f(E_1+\eta
\omega)[\Delta_0^s]^{2}\delta(E_1-E_2+\eta\omega)}{\mathcal{D}^s(E_1)(\mathcal{D}^s(E_1)+\eta\omega)^\ast}.
\end{eqnarray}
Thus, the d.c. current generated by the time-varying parameters
has the following final expression:
\begin{eqnarray}
\langle
I_l^{s(1)}(t)\rangle&=&\frac{e\tilde{\mathcal{Q}}^s_1\mathcal{Q}^s_1\Delta_0^s[1-\Delta_0^s]}{h}\int
\frac{f(E)dE}{|\mathcal{D}^s(E)|^2}+\nonumber \\
&+&\frac{2e\mathcal{Q}^s_3J_1(2\pi\Phi_R^{\omega})}{h}\times
\nonumber \\
\label{eq:dc-curr-u=0} &\times& Re\{\sum_{\eta=\pm 1
}\int \frac{\omega
\mathcal{R}^{s}_{\eta}f(E)dE}{\mathcal{D}^s(E)(\mathcal{D}^s(E)+\eta\omega)}\}.
\end{eqnarray}
Here $\tilde{\mathcal{Q}}^s_1$ is a coefficient obtained setting
$\gamma=0$ in $\mathcal{Q}^s_1$ (for the left lead). The current
(\ref{eq:dc-curr-u=0}) contains terms proportional to
$\varepsilon_\omega^2$ and $\left(\Phi_R^{\omega}\right)^2$ that
can be interpreted as rectification terms, and terms proportional
to $\varepsilon_{\omega}\Phi_R^{\omega}$ which contain information
on the non-adiabatic pumping process, as will be clear below.

\subsection{Weakly interacting case ($U \approx 0$)}
To perform the analysis in the weakly interacting case, we
consider  as negligible the terms proportional to $U\omega$. In
addition we also consider $U/\omega $ as a small quantity, being
$U$ of the same order of $\varepsilon_\omega$, $\Phi_R^\omega$.
Within the Hartree-Fock theory we need to determine the energy of
the quantum dot $\varepsilon_\sigma(t)=\varepsilon_0+U
n_{\bar{\sigma}}(t)$ with $n_{\bar{\sigma}}(t)\equiv\langle
n_{\bar{\sigma}}(t)\rangle$. By using the single photon
approximation, we write the occupation number as a trigonometric
series:
\begin{eqnarray}
n_{\sigma}(t)=a_\sigma^{(0)}+a_\sigma^{(1)}\sin(\omega t
)+a_\sigma^{(2)}\cos(\omega t),
\end{eqnarray}
where the unknown coefficients $a_\sigma^{(i)}$ have to be
determined self-consistently. In the interacting case the retarded
Green's function takes the following form:
\begin{eqnarray}
\label{eq:gr_app}
G^{r(1)}_{sp}(t,t')&=&-i\delta_{sp}\theta(t-t')\times \nonumber \\
&\times&\exp\Bigl\{-i[\varepsilon_0+U a_{\bar{s}}^{(0)}
-i\mathcal{Q}^{s}_1/2)](t-t')\Bigl\}\times \nonumber \\
&\times&\exp\Bigl\{-i[\lambda_1^s \int_{t'}^{t}dt_1\sin(\omega
t_1)+\nonumber \\ &+&\lambda_2^s \int_{t'}^{t}dt_1\cos(\omega
t_1)] \Bigl\},
\end{eqnarray}
where the coefficients $\lambda_i^s$ have been defined as follows:
\begin{eqnarray}
\lambda_1^s&=&
\varepsilon_{\omega}\cos(\varphi)+iJ_1(2\pi\Phi_R^{\omega})\mathcal{Q}_3^s+U
a_{\bar{s}}^{(1)}\nonumber \\
\lambda_2^s&=&\varepsilon_{\omega}\sin(\varphi)+U
a_{\bar{s}}^{(2)}.
\end{eqnarray}
Note that since the coefficients $a^{(i)}_s$ appear as a factor of
the interaction $U$, we have to calculate them only up to the zero
order approximation in $U$ and $\omega$.
From the lesser Green's function obtained by (\ref{eq:gr_app}), we
can write the occupation number  in the following form:
\begin{eqnarray}
n_{\sigma}(t)&\approx&
\frac{(\Delta_0^{\sigma})^2}{2\pi}[\mathcal{Q}_1^{\sigma}-2\mathcal{Q}_3^{\sigma}J_1(2\pi\Phi_R^{\omega})\sin(\omega
t )]\times \nonumber \\ &\times&\int\frac{f(E)dE}{
|\mathcal{D}_0^s(E)|^2 },
\end{eqnarray}
where
$\mathcal{D}_0^{\sigma}(E)=E-\varepsilon_0+i\mathcal{Q}_1^s/2$. By
comparing the above expression with the
$G_{\sigma\sigma}^{<(1)}(t,t)$ obtained by using (\ref{eq:gr_app})
one gets the following set of self-consistency equations:
\begin{eqnarray}
&&a_{\sigma}^{(0)}=\frac{\mathcal{Q}_1^{\sigma}}{2\pi}\int\frac{f(E)dE}{
|\mathcal{D}_0^{\sigma}(E)|^2 }\nonumber \\
&&a_{\sigma}^{(1)}=-2\frac{\mathcal{Q}_3^{\sigma}}{\mathcal{Q}_1^{\sigma}}J_1(2\pi\Phi_R^{\omega})a_{\sigma}^{(0)}\nonumber
\\ &&a_{\sigma}^{(2)}=0.
\end{eqnarray}
Once the above equations are solved, the d.c. current can be
written as in Eq.(\ref{eq:dc-curr-u=0}) with the following
interaction-induced shift:
$\varepsilon_0\longrightarrow\varepsilon_0+U
a_{\bar{\sigma}}^{(0)}$, $\Lambda_1^s\longrightarrow\Lambda_1^s +U
a_{\bar{\sigma}}^{(1)}$.

\subsection{The spin and charge currents}\label{sec:zero-temp-curr}

To obtain an analytical expression of the d.c. current pumped in
the left lead in the presence of a weak interaction and zero
temperature, we need an approximated expression for
$\mathcal{R}^s_{\eta}$ in (\ref{eq:dc-curr-u=0}) in the limit of
$\frac{\varepsilon_\omega}{\omega}\ll 1$, and $2\pi\Phi_R^\omega
\ll 1$. In this limit one can use $J_1(x)\simeq x/2$,
$I_1(x)\simeq x/2$, $J_0(x)\simeq 1$, $I_0(x)\simeq 1$ (see
APPENDIX \ref{app:bessel-approx}) and thus the coefficients
$\mathcal{R}^s_{\eta}$ take the following simplified form:
\begin{eqnarray}
\mathcal{R}^s_{\eta}\approx\frac{\pi \Phi_R^\omega
\mathcal{Q}_3^s}{2\omega}+\eta \frac{\varepsilon_\omega
\sin(\varphi)}{2 \omega}-i\frac{(\varepsilon_\omega\cos(\varphi)+U
a^{(1)}_{\bar{s}} )}{2\omega},
\end{eqnarray}
while the quantity $\Delta_0^s[1-(\Delta_0^s)]$ can be written as
follows:
\begin{eqnarray}
\Delta_0^s[1-\Delta_0^s]&\approx&\frac{1}{2}\Bigl[
\frac{\varepsilon_\omega^2}{\omega^2}- \frac{(\pi \Phi_R^\omega
\mathcal{Q}_3^s)^2}{\omega^2}+\nonumber \\
&+&2\frac{a_{\bar{s}}^{(1)}U\varepsilon_\omega\cos(\varphi)}{\omega^2}\Bigl]+\mathcal{O}(1/\omega^4).
\end{eqnarray}
Plugging these expressions in (\ref{eq:dc-curr-u=0}) and
performing the integral over the frequency,
after an expansion up to the second order in $\omega$, we can
write the d.c. current in the single photon approximation (in
units of $2\Gamma^0e/\hbar$) as follows:
\begin{eqnarray}\label{eq:current-final-zero-temp}
\langle i^{(1)}_\sigma \rangle
&=&\frac{\tilde{q}_1^{\sigma}a_{\sigma}^{(0)}}{2\omega^2}\Bigl[\varepsilon_{\omega}^2-(2\pi\Phi_R^\omega
q_3^{\sigma})^2+2a_{\bar{\sigma}}^{(1)}U\varepsilon_\omega
\cos(\varphi)\Bigl]+\nonumber \\ &-&\pi (q_3^\sigma
\Phi_R^\omega)^2\Bigl[\frac{\mu-\varepsilon_0+U(a^{(0)}_{\sigma}-a^{(0)}_{\bar{\sigma}})}{|\mathcal{D}^{\sigma}(\mu)|^2}\Bigl]+\nonumber
\\
&+&q_3^{\sigma}\frac{\varepsilon_\omega\Phi_R^\omega}{2|\mathcal{D}^{\sigma}(\mu)|^2}\Bigl\{\frac{\omega\sin(\varphi)}{|\mathcal{D}^{\sigma}(\mu)|^2}[(\mu-\varepsilon_0-U
a_{\bar{\sigma}}^{(0)})^2-(q_1^{\sigma})^2]\nonumber \\
&+&2q_1^{\sigma}\cos(\varphi)\Bigl\},
\end{eqnarray}
where the energies are measured in units of $\Gamma^0$, while we
defined $q_i^\sigma\equiv\mathcal{Q}_i^{\sigma}/(2\Gamma^0)$. The
non-dimensional charge and spin currents, i.e. $I_c$ and $I_s$,
can be defined as $I_c=\sum_\sigma i_\sigma$ and $I_s=\sum_\sigma
\sigma i_\sigma$. The main feature of the expression for the d.c.
current is the presence of a non-sinusoidal current-phase relation
already in weak-pumping. Indeed, contrarily to the adiabatic case
characterized by a current-phase relation with definite odd parity
(i.e. $I_c(-\varphi)=-I_c(\varphi)$ ) in the time-dependent case
any parity with respect to the sign reversal of $\varphi$ is
expected in the pumped current. This behavior is mainly related to
finite frequency effects as well as to interaction effects.
Eq.(\ref{eq:current-final-zero-temp}) represents the main result
of this work.

\section{Numerical results and discussion}\label{sec:results}
In order to make a comparison with the available experimental
data, we set $\Gamma^0 \sim 10 \mu$eV[\onlinecite{dicarlo-exp03}].
Such quantity is related to the dwell time $\tau_d$ by the
following relation $E_\tau=h/\tau_d \sim 2 \Gamma^0$. Such
quantity is relevant to define the various transport regimes at
varying frequency $\omega$. Indeed, for value of $\omega \tau_d
\ll 1$ one deals with the adiabatic regime, while in the opposite
limit, i.e. $\omega \tau_d \gg 1 $, the non-adiabatic regime is
approached. For typical experimental frequencies ranging from 10
MHz up to 20 GHz, $\omega \tau_d$ varies from $\sim 10^{-2}$ up to
order 10
and thus the MHz range of frequency can be safely considered as
adiabatic. The adimensional frequency $\omega$ which appears in
Eq.(\ref{eq:current-final-zero-temp}) is defined as $\omega\equiv
\hbar\omega/\Gamma^0=\omega\tau_d/\pi$.
In this way a frequency of 25 MHz corresponds to $\omega=0.01$,
100 MHz to $\omega=0.04$, 1 GHz to $\omega=0.4$.
In the following we study the behavior of charge and spin currents
in the range of frequency $\omega \in [0.1,0.5]$, thus our
analysis is valid from adiabatic up to the moderate non-adiabatic
limit. We also set the chemical potential $\mu$ as the zero of
energy. From the analysis of the current $i_\sigma$, we notice the
presence of two classes of terms contributing to the currents: 1)
Terms proportional to $\varepsilon_\omega^2$ or
$\left(\Phi_R^\omega \right)^2$; 2) terms proportional to
$\Phi_R^\omega \varepsilon_\omega$. The first type of terms are
non-adiabatic in nature. The second class of terms contains a term
proportional to $\omega$, which can be recognized as the quantum
pumping contribution, and a frequency independent term
proportional to $\cos(\varphi)$ which can be interpreted as a
rectification contribution. Such term is responsible for the
non-sinusoidal behavior that leads to an anomalous current-phase
relation as observed in Ref.[\onlinecite{dicarlo-exp03}] (page 3,
first column, line 2). Very interestingly, the interaction effects
also lead to a cosine term which is proportional to $U
\varepsilon_\omega \Phi_R^{\omega}\cos(\varphi)/\omega^2$. Such a
term produces a deviation from the sinusoidal behavior also for
small values of the energy $U$. Finally, the current $i_\sigma$
vanishes when the amplitude of the modulation $\varepsilon_\omega$
and $\Phi_R^\omega$ go simultaneously to
zero.\\

\begin{figure}[htbp]
\centering
\includegraphics[scale=.55]{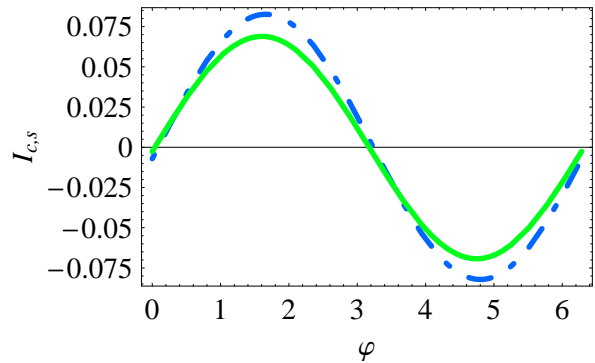}
\caption{Currents of charge (dashed-dotted line) and spin (full
line) as a function of $\varphi$ obtained for the following choice
of parameters: $\gamma=0.05$, $\Phi_{AB}=0.49$, $\Phi_R^0=0.02$,
$\Phi_R^\omega=0.01$, $\varepsilon_0=0$,
$\varepsilon_\omega=0.025$, $\omega=0.1$ and $U=0$.}
\label{fig:current-phase}
\end{figure}

\begin{figure}[htbp]
\centering
\includegraphics[scale=.55]{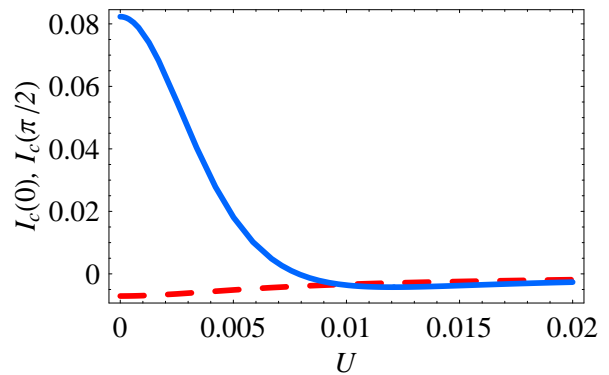}
\caption{Currents of charge computed at $\varphi=0$,
$I_c(\varphi=0)$ (dashed line), and $\varphi=\pi/2$,
$I_c(\varphi=\pi/2)$ (full line), as a function of $U$ obtained
for the following choice of parameters: $\gamma=0.05$,
$\Phi_{AB}=0.49$, $\Phi_R^0=0.02$, $\Phi_R^\omega=0.01$,
$\varepsilon_0=0$, $\varepsilon_\omega=0.025$, $\omega=0.1$.}
\label{fig:current-u}
\end{figure}

\begin{figure}[htbp]
\centering
\includegraphics[scale=.55]{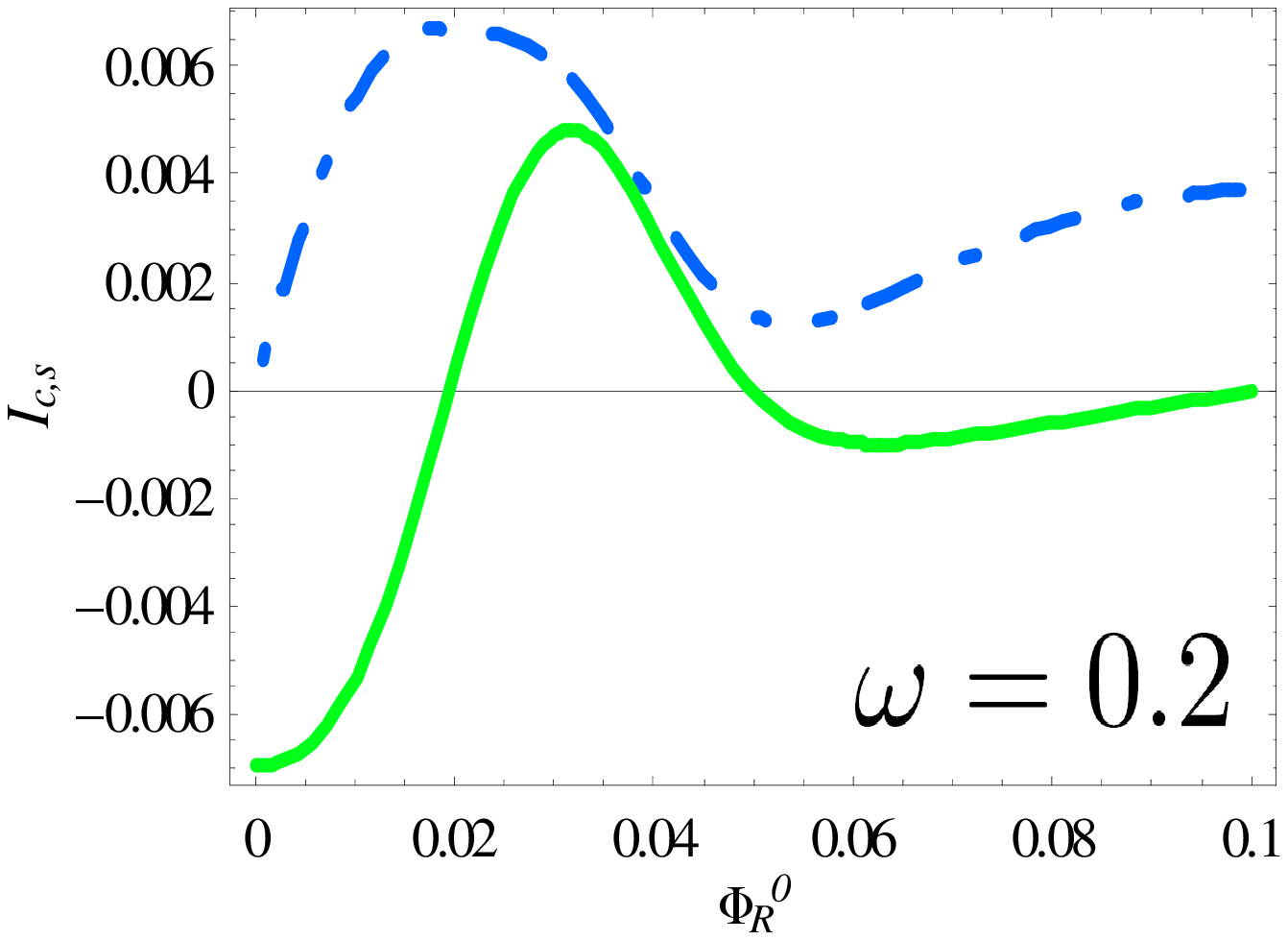}\\
\includegraphics[scale=.55]{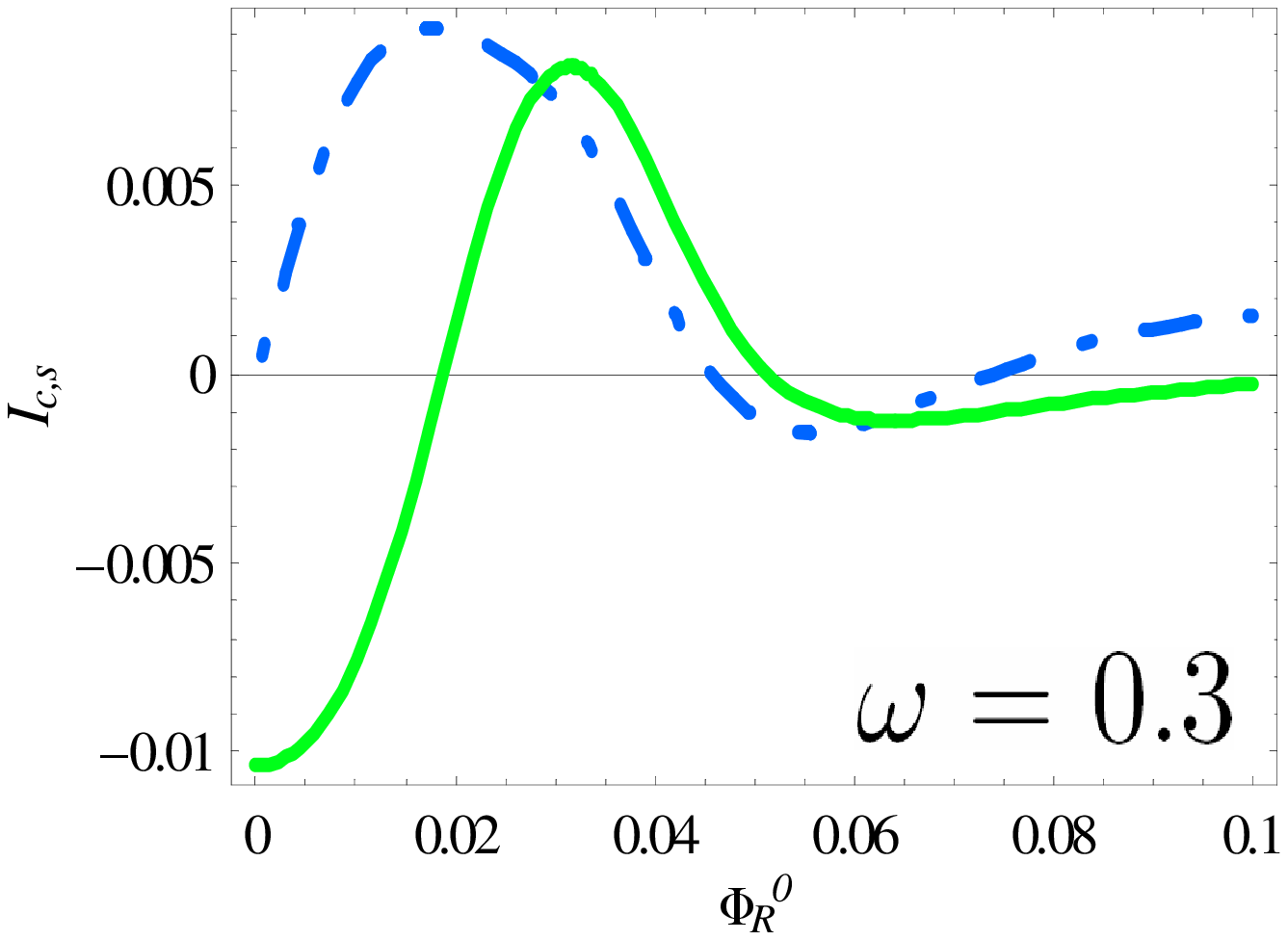}\\
\includegraphics[scale=.55]{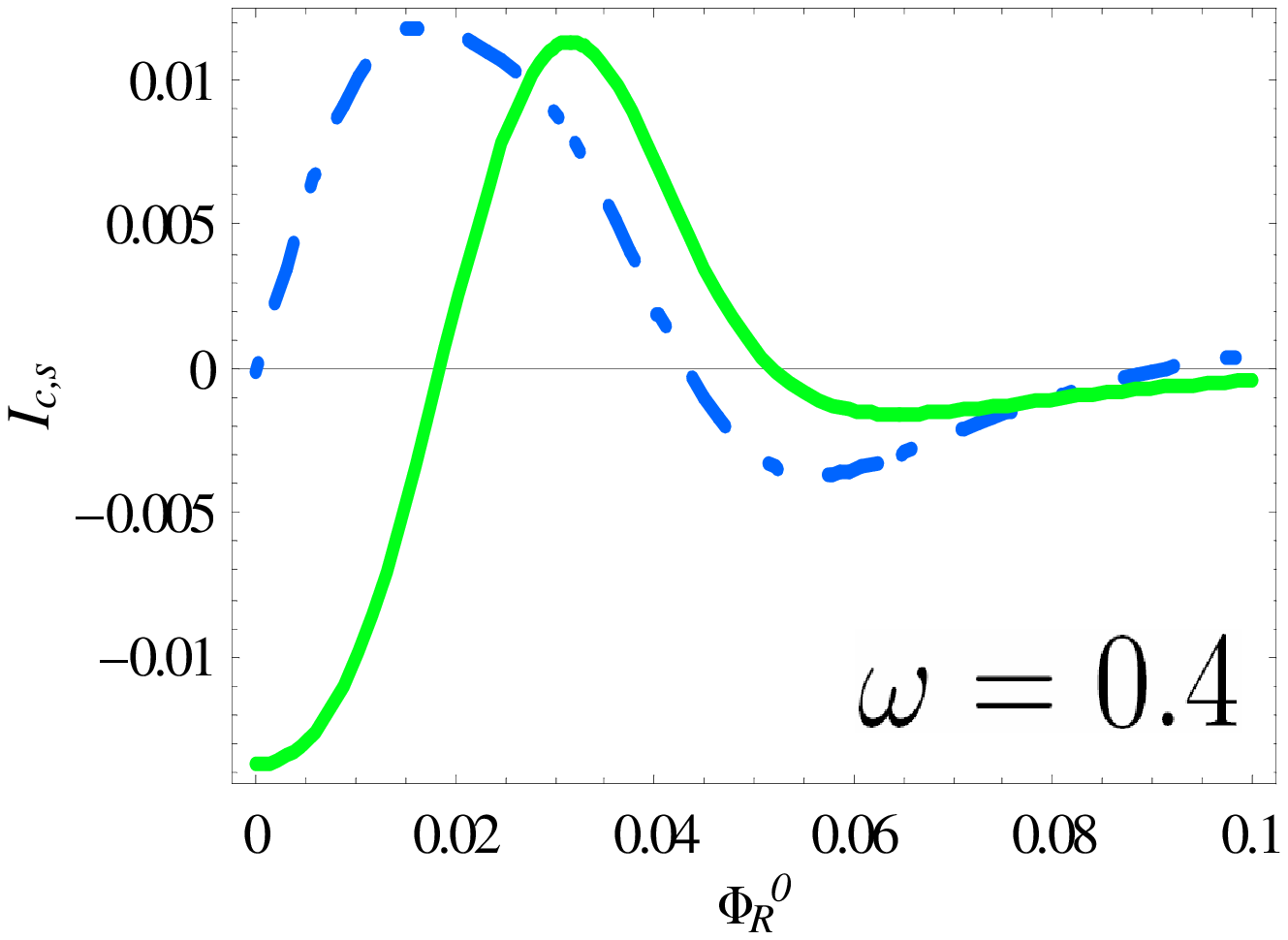}
\caption{Charge (dashed-dotted line) and spin (full line) currents
as a function of $\Phi_R^0$ obtained for the following choice of
parameters: $\gamma=0.05$, $\Phi_{AB}=0.49$, $\Phi_R^\omega=0.01$,
$\varepsilon_0=-0.025$, $\varepsilon_\omega=0.05$,
$\varphi=5\pi/4$, $U=0$. The upper panel is obtained for
$\omega=0.2$, the middle panel for $\omega=0.3$ and the lower
panel for $\omega=0.4$.} \label{fig:current-fr0}
\end{figure}

\begin{figure}[htbp]
\centering
\includegraphics[scale=.55]{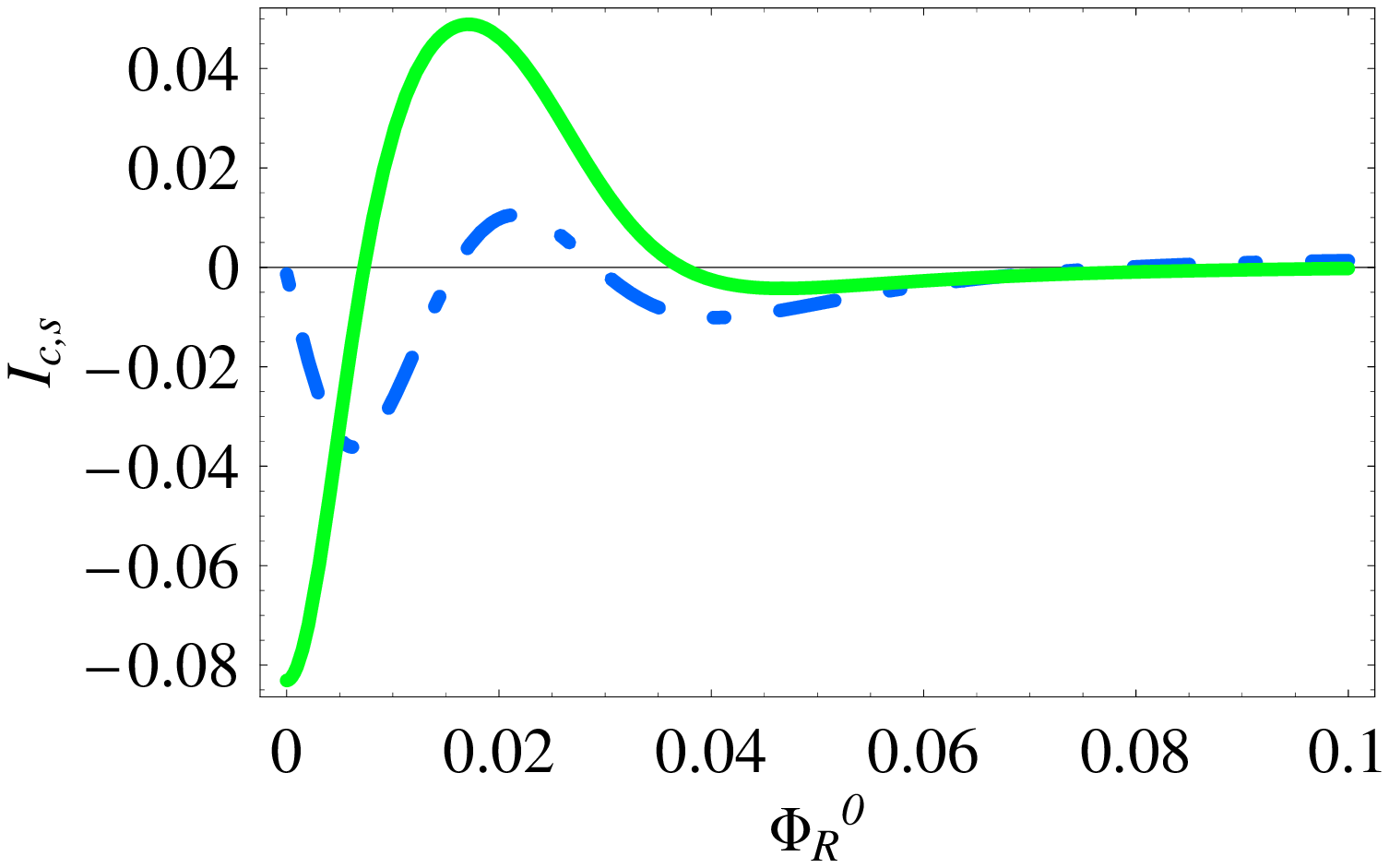}\\
\includegraphics[scale=.55]{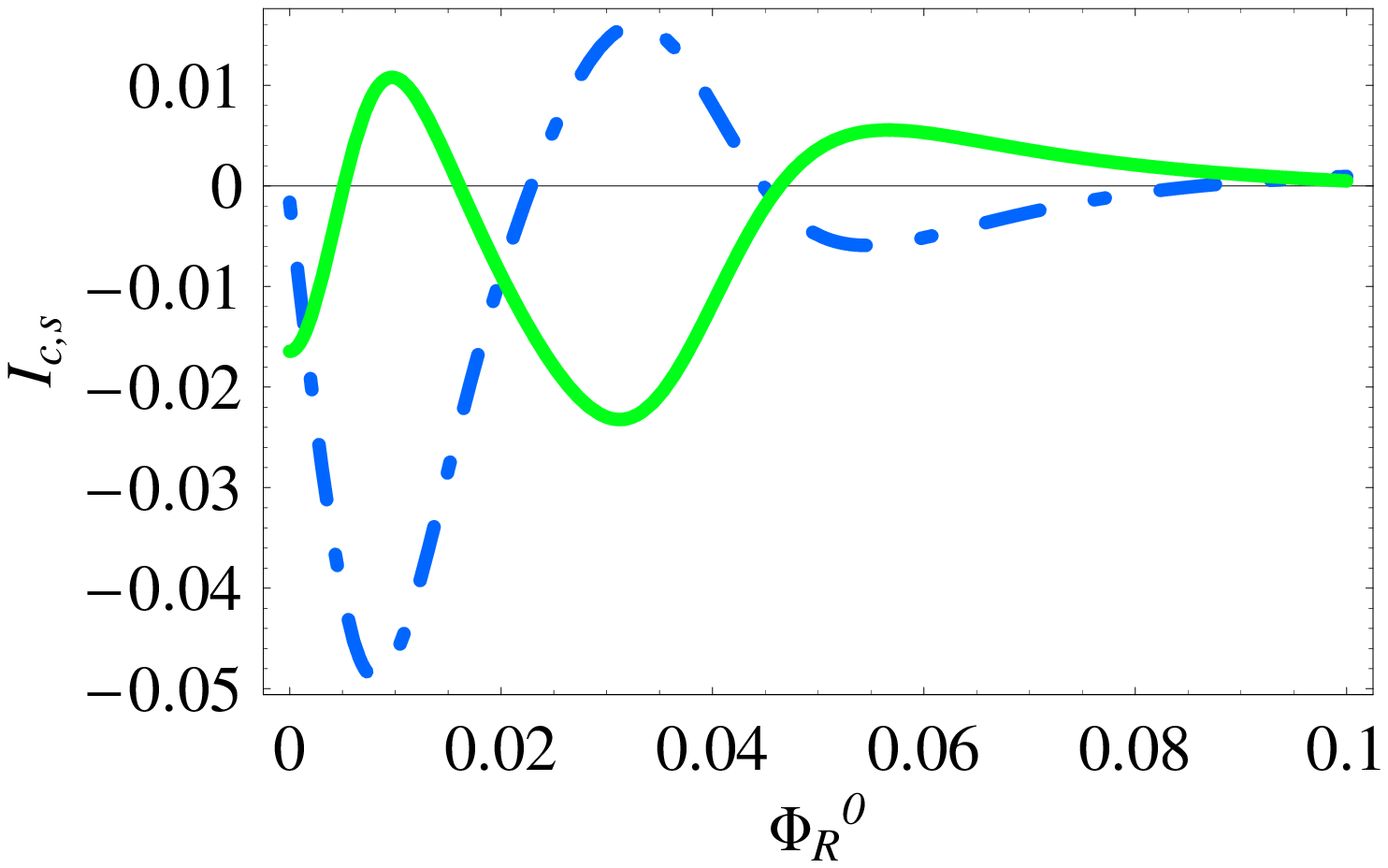}
\caption{Charge (dashed-dotted line) and spin (full line) currents
as a function of $\Phi_R^0$ obtained for the following choice of
parameters: $\gamma=0.05$, $\Phi_R^\omega=0.01$,
$\varepsilon_0=-0.025$, $\varepsilon_\omega=0.05$,
$\varphi=5\pi/4$, $\omega=0.3$, $U=0.02$ and $\Phi_{AB}=0.49$ in
the upper panel, $\Phi_{AB}=0.52$ in the lower panel.}
\label{fig:current-fr0-u}
\end{figure}

In Fig.\ref{fig:current-phase} the charge (dashed-dotted line) and
spin (full line) currents, namely $I_c$ and $I_s$, as a function
of the phase difference $\varphi$ between the time-varying
parameters are reported for the following choice of parameters:
$\gamma=0.05$, $\Phi_{AB}=0.49$, $\Phi_R^0=0.02$,
$\Phi_R^\omega=0.01$, $\varepsilon_0=0$,
$\varepsilon_\omega=0.025$, $\omega=0.1$ and $U=0$. A
sinusoidal-like behavior is observed even though the charge pumped
for $\varphi=0$ is different from zero and of the order $10^{-3}$.
This is a fingerprint of the anomalous current-phase relation, as
discussed above.
To put in evidence the dependence on the interaction $U$, we
present in Fig.\ref{fig:current-u} the charge current computed at
$\varphi=0$ (dashed line) and $\varphi=\pi/2$ (full line) as a
function of $U$ taking the remaining parameters as in
Fig.\ref{fig:current-phase}. Smaller values of the interaction
favours deviation from the sinusoidal behavior.
Below we concentrate on the role of spin-orbit interaction and
choose the Aharonov-Bohm flux close to half integer values in unit
of the flux quantum $\Phi_0=h/e$ where the charge current is
activated by photon-assisted tunneling (PAT).
 Away from the above values of the Aharonov-Bohm flux  the currents present an oscillating behavior as a function of
the applied magnetic flux $\Phi_{AB}$ similar to the one already
discussed in a previous work\cite{ac-pump-citro06}.\\
In Fig.\ref{fig:current-fr0} we plot $I_c$ (dashed-dotted line)
and $I_s$ (full line) as a function of the static Aharonov-Casher
phase $\Phi_R^0$ for pumping frequency $\omega=0.2$, $\omega=0.3$,
$\omega=0.4$ (from top to bottom) and by setting the remaining
parameters as follows: $\gamma=0.05$, $\Phi_{AB}=0.49$,
$\Phi_R^\omega=0.01$, $\varepsilon_0=-0.025$,
$\varepsilon_\omega=0.05$, $\varphi=5\pi/4$, $U=0$.
By increasing the pumping frequency from $\omega=0.2$ (500 MHz) up
to $0.4$ (1 GHz) zeros of the charge currents start to appear and
thus it is possible to obtain pure spin currents in the
non-adiabatic regime. It is worth mentioning that currents of
amplitude $10^{-2}$ in dimensionless units correspond to $\sim
50$pA in
dimensional unit, thus the pure spin current we find is sizable.\\
To analyze the role of a weak Coulomb interaction, we plot in the
upper panel of Fig.\ref{fig:current-fr0-u} the charge and spin
currents for the same parameters as in Fig.\ref{fig:current-fr0}
($\omega=0.3$) and by setting $U=0.02$. A qualitatively different
behavior of the charge current as a function of the
Aharonov-Casher flux is observed.
In particular, when the interaction energy $U$ is of the same
order of magnitude of the pumping frequency $\omega$, additional
zeros of the charge current appear and this is a very appealing
situation for spintronics devices.  For instance, looking at
Fig.\ref{fig:current-fr0-u} (upper panel), one observes a pure
spin current close to $\Phi_R^0=0.015$ and $0.03$.\\
In the lower panel of Fig.\ref{fig:current-fr0-u} we plot charge
and spin currents as done in the upper panel and by setting the
Aharonov-Bohm flux to $\Phi_{AB}=0.52$. In this case, a
characteristic oscillating behavior of the currents controlled by
using a magnetic flux is visible.

\begin{figure}[htbp]
\centering
\includegraphics[scale=.45]{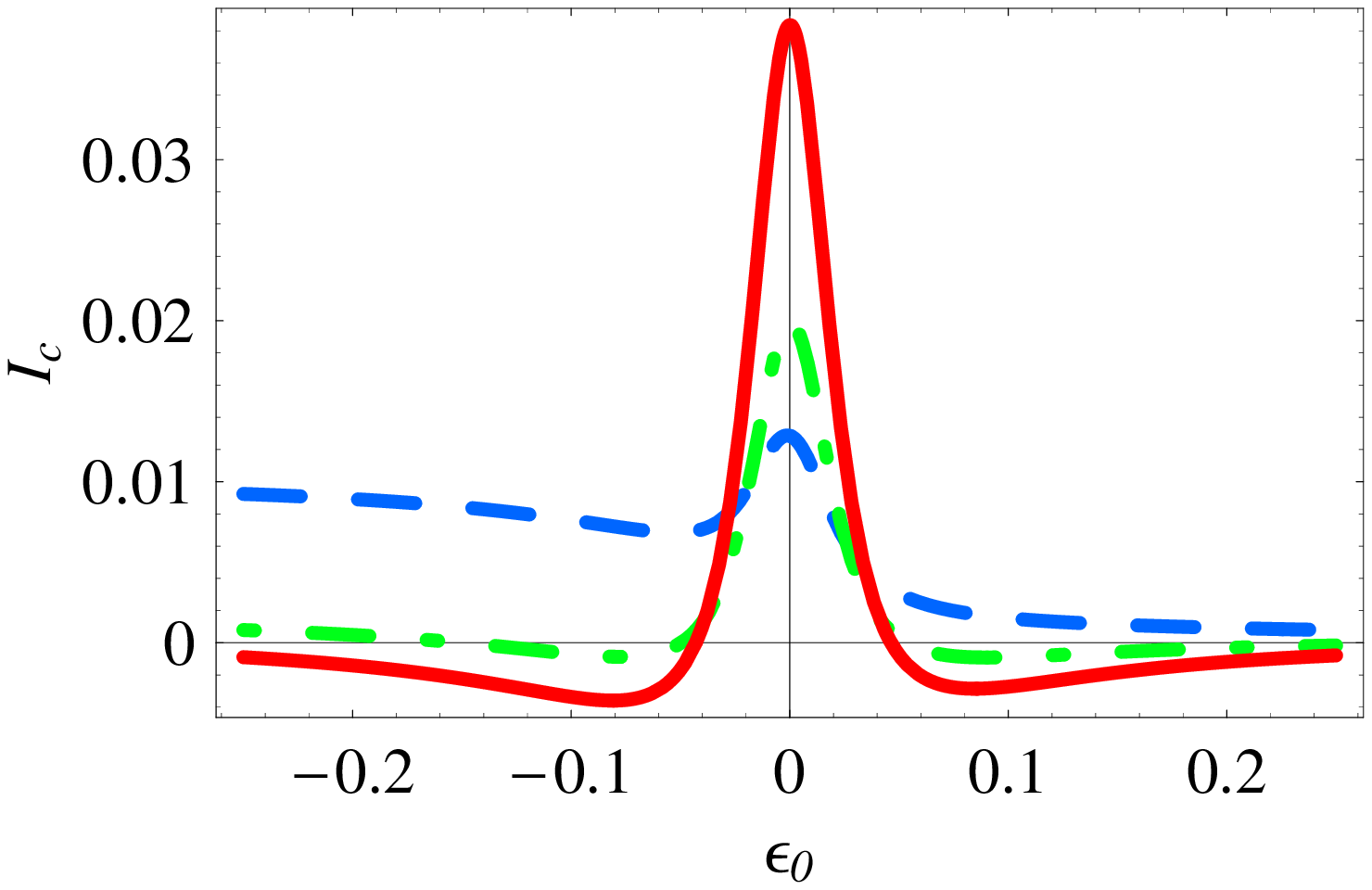}\\
\includegraphics[scale=.45]{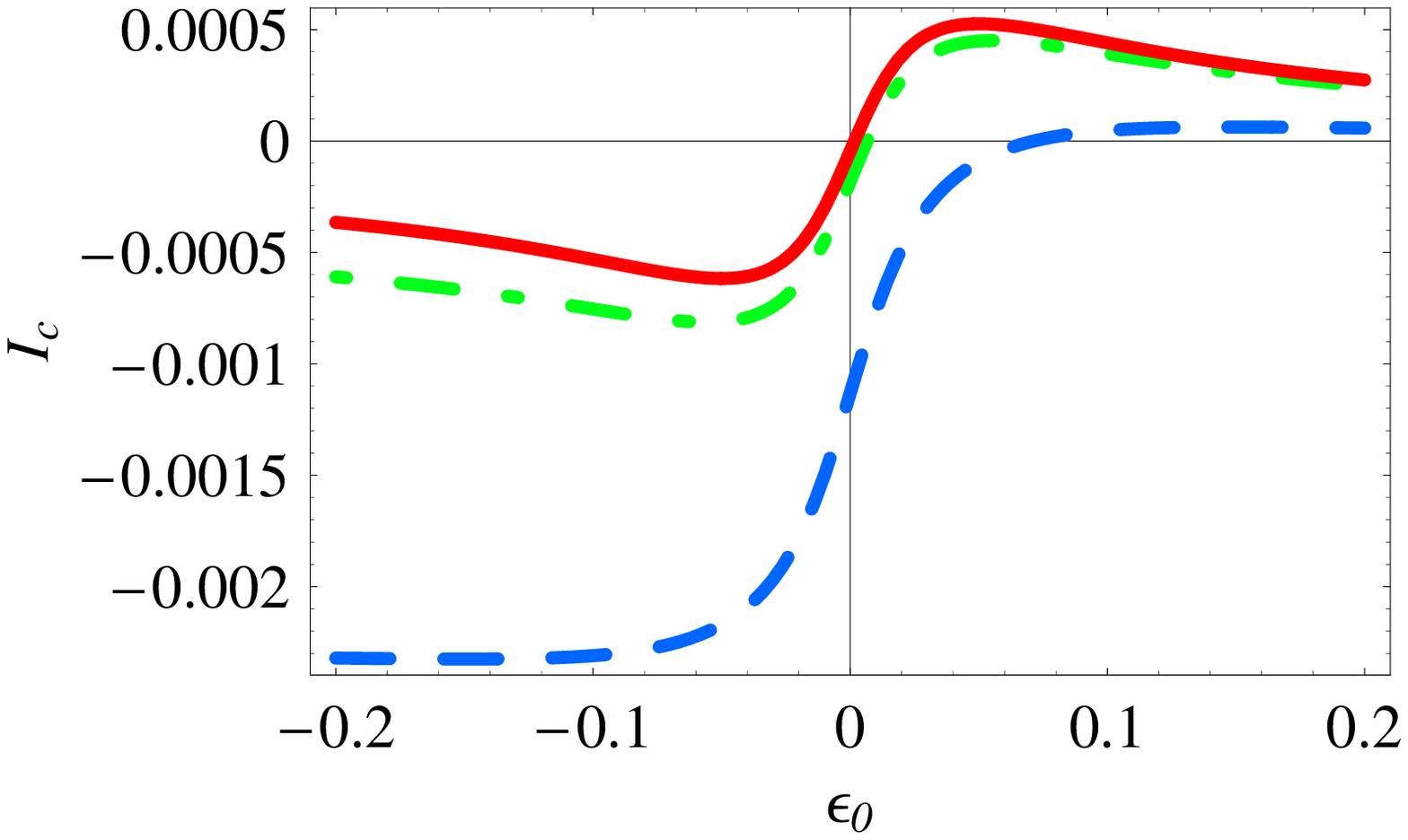}
\caption{Charge current $I_c$ as a function of $\varepsilon_0$
obtained for the following choice of parameters: $\gamma=0.05$,
$\Phi_{AB}=0.49$, $\Phi_R^0=0.05$, $\Phi_R^\omega=0.01$,
$\varphi=\pi/2$, $U=0$ and $\varepsilon_\omega=0.05$ for upper
panel, $\varepsilon_\omega=0$ for lower panel. Each panel contains
curves obtained for $\omega=0.1$ (dashed line), $\omega=0.25$
(dashed-dotted line), $\omega=0.5$ (full line).}
\label{fig:current-e0}
\end{figure}

\begin{figure}[htbp]
\centering
\includegraphics[scale=.45]{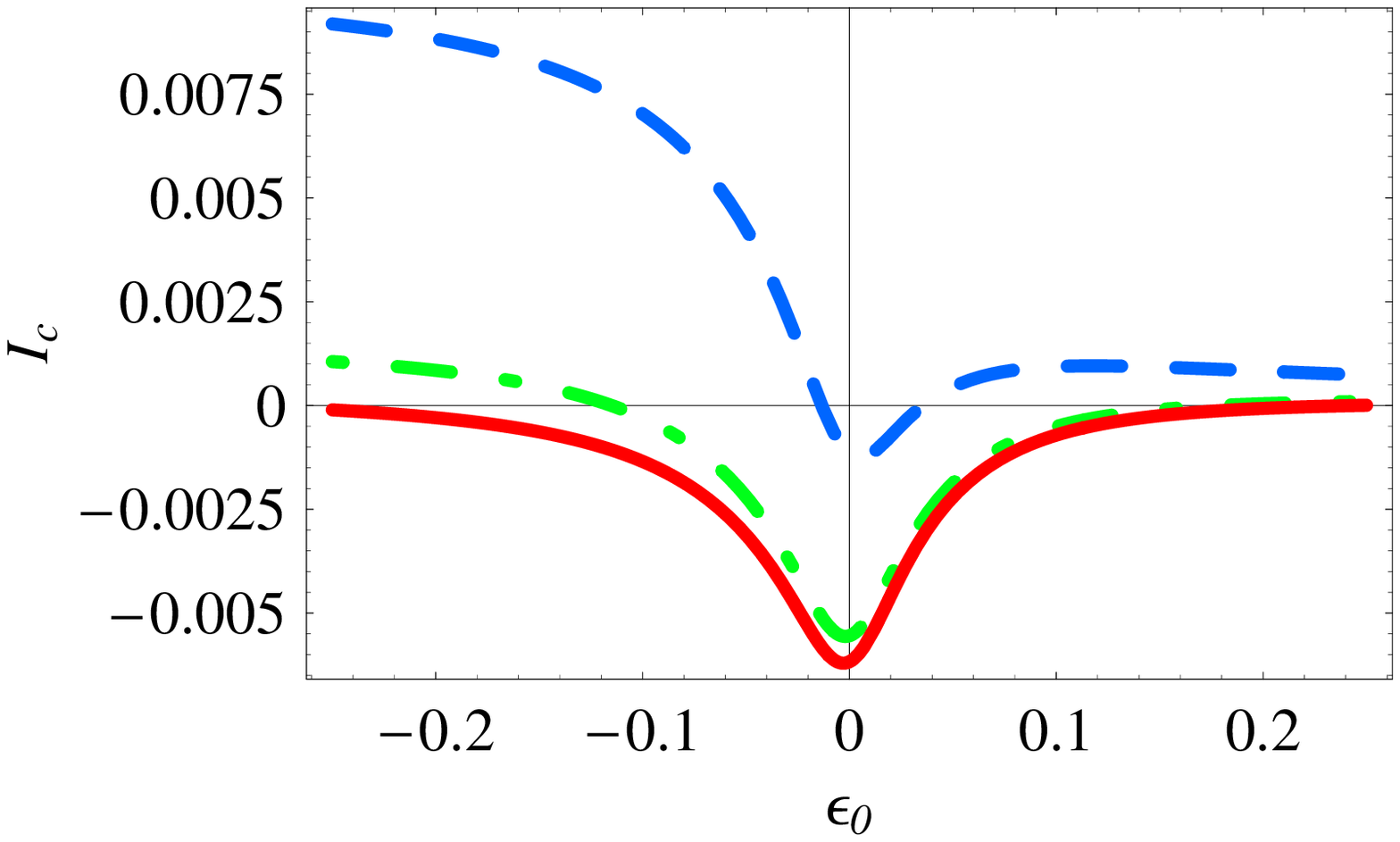}\\
\includegraphics[scale=.45]{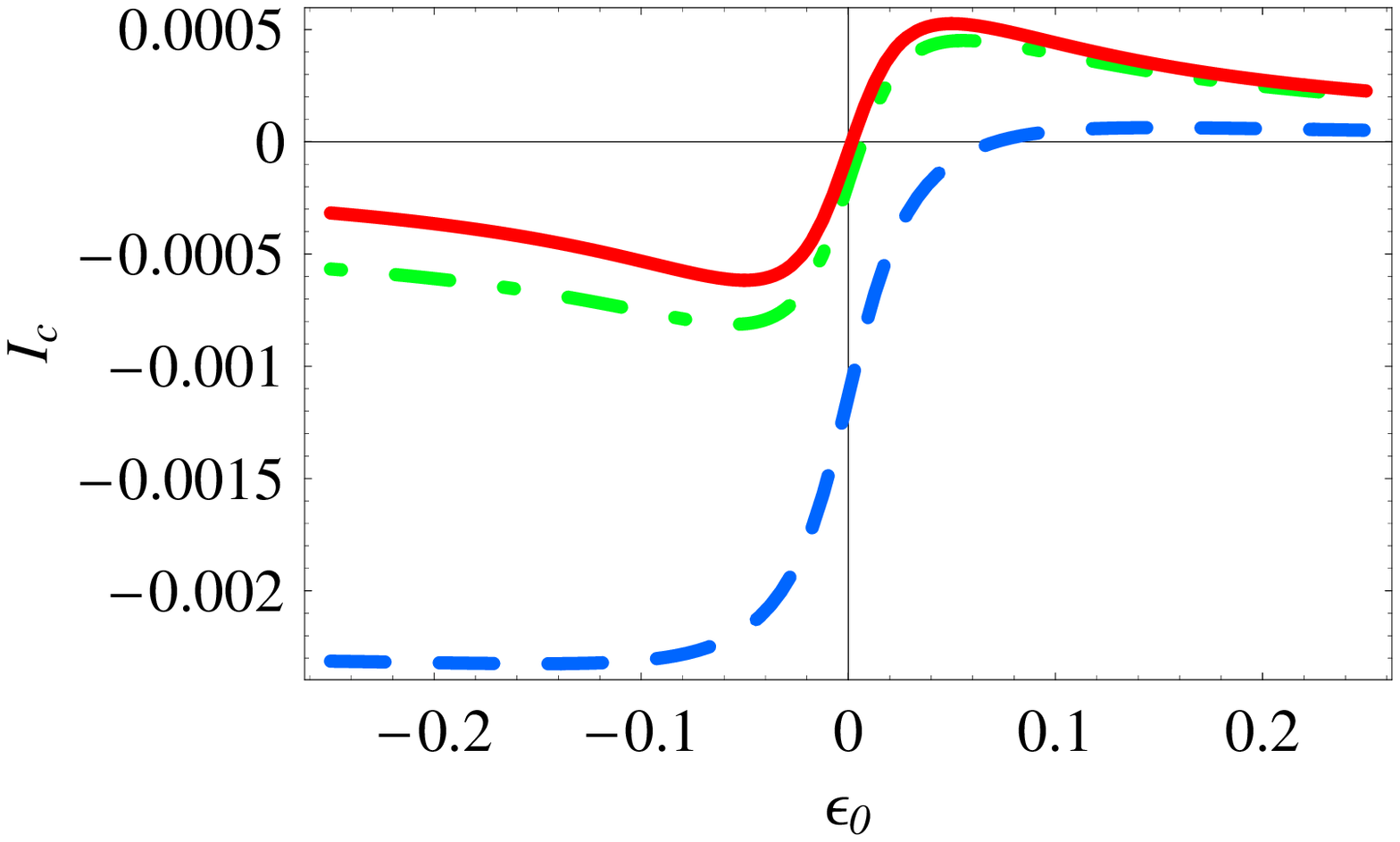}
\caption{Charge current $I_c$ as a function of $\varepsilon_0$
obtained for the following choice of parameters: $\gamma=0.05$,
$\Phi_{AB}=0.49$, $\Phi_R^0=0.05$, $\Phi_R^\omega=0.01$,
$\varphi=0$, $U=0$ and $\varepsilon_\omega=0.05$ for upper panel,
$\varepsilon_\omega=0$ for lower panel. Each panel contains curves
obtained for $\omega=0.1$ (dashed line), $\omega=0.25$
(dashed-dotted line), $\omega=0.5$ (full line).}
\label{fig:current-e0-phi0}
\end{figure}

Another interesting phenomenon is the asymmetric contribution to
the current of the photon absorption and emission as a function of
the dot level $\varepsilon_0$, as also reported in
Ref.[\onlinecite{non-adiabatic-pump-braun08}]. When the dot level
lies above the Fermi level ($\varepsilon_0>0$), an electron on the
dot can jump in the left lead by emitting a photon. For
$\varepsilon_0<0$, an electron on the dot can reach the left lead
only by means of the absorption of a photon since no voltage bias
or temperature gradient are present. Because of the interference
between these two-photon sources, boson-assisted tunneling onto
the dot gets suppressed while tunneling out of the quantum dot is
enhanced. The asymmetric behavior of the d.c. current as a
function of the dot level is shown in Fig.\ref{fig:current-e0}.
In the upper panel, we plot the charge current $I_c$ as a function
of the dot level $\varepsilon_0$ and by setting the remaining
parameters as: $\gamma=0.05$, $\Phi_{AB}=0.49$, $\Phi_R^0=0.05$,
$\Phi_R^\omega=0.01$, $\varphi=\pi/2$, $U=0$ and
$\varepsilon_\omega=0.05$. As can be seen, when the frequency
$\omega$ is increased from 0.1 (dashed line) up to 0.5 (full line)
a strong peak is formed at Fermi energy and the asymmetry of the
current with respect to the $\varepsilon_0=0$ becomes more
evident. It is worth to mention that, since the relative phase
$\varphi$ is $\pi/2$, all the terms in the current proportional to
$\cos(\varphi)$ are suppressed, while the pumping term takes its
maximum value. In the lower panel we set $\varepsilon_\omega=0$,
while the remaining parameters are fixed as in the upper panel. In
this case the device works as a single-parameter pump associated
to the Aharonov-Casher flux and the current is proportional to
$\left(\Phi_R^{\omega}\right)^2$. A comparison between the upper
and the lower panel shows that the pumping mechanism is the
dominant one at the Fermi energy. Furthermore, we have verified
that a small interaction does not
alter too much the picture given so far.\\
The same analysis performed in Figs.\ref{fig:current-e0} can be
repeated by setting $\varphi=0$ to include the  $\cos(\varphi)$
contribution. In the upper panel of Fig.\ref{fig:current-e0-phi0}
we plot the current obtained for $\varepsilon_\omega=0.05$, while
in the lower panel this parameter is set to zero (single-parameter
pump).
By comparing the results, one observes an enhancement of the
absolute value of the high frequency currents in the case of
double-parameter modulation (upper panel) and close to the
Fermi energy.\\
From the analysis above one observes that, within the considered
parameters region, the dominant mechanism for the generation of
the d.c. current is the finite frequency quantum pumping. Indeed,
close to the Fermi energy such currents take values which range
from $\sim 70$pA up to $\sim 190$pA (see the upper panel of
Figs.\ref{fig:current-e0}), while in the other cases the generated
currents present values of about $10 \%$ of those induced by the
pumping process. Thus, for $\Phi_{AB}$ close to half-integer
values the quantum pumping induces the main contribution to the
current, while away from this flux region the rectification
currents are dominant.

\section{Conclusions}\label{sec:conclusions}
We studied the time-dependent charge and spin transport (pumping)
in a Aharonov-Bohm-Casher ring sequentially coupled to a weakly
interacting quantum dot by using a non-equilibrium Green's
function approach. By varying a considerable number of parameters,
we showed that the proposed device can work as a spin current
generator and analyzed all its characteristics, including
rectification effects. When the energy level $\varepsilon(t)$ on
the dot and the Aharonov-Casher flux are periodically modulated in
time with a frequency $\omega$, a d.c. current is observed in the
leads. Contrarily to the adiabatic case, the current-phase
relation presents two additional cosine terms: The first one comes
from the interaction on the dot, while the second can be
interpreted as a rectification effect, as already noted in
Ref.[\onlinecite{dicarlo-exp03}]. We also showed that Coulomb
interaction effects can enhance the rectification effects. As a
function of the spin-orbit interaction and close to the
non-adiabatic regime, the results of the charge current show the
appearance of additional zeros at varying the Aharonov-Casher
flux. Thus, the finite frequency regime close to 750 MHz
($\omega=0.3$) is suitable to obtain pure spin currents useful in
spintronics. Such currents are of the order of magnitude of $\sim
100$pA as detected in the experiments on quantum
dots\cite{watson-exp03}. Finally, the analysis as a function of
the dot level showed a characteristic asymmetric behavior and the
comparison between the single parameter pump and double-parameters
one showed a considerable increase of the d.c. current in the
second case.  The proposed device can be easily fabricated on a
GaAs/AlGaAs two-dimensional electron gas using e-beam lithography
to define the ring and dot region modifying, for instance, the
system studied in Ref.[\onlinecite{watson-exp03}].

\section*{ACKNOWLEDGMENTS}
One of the authors (F. R.) would like to honor the memory of
Antonio Calderaro who prematurely terminated his human adventure
when the authors were writing this work.

\appendix{}
\section{Bessel expansion}\label{app:bessel-expansion}
Throughout the paper the following expansions have been exploited:
\begin{eqnarray}
\sin(x \sin(\theta))&=&
2\sum_{n=1}^{\infty}J_{2n-1}(x)\sin((2n-1)\theta)\nonumber \\
\cos(x \sin(\theta))&=& J_0(x)+2\sum_{n=1}^{\infty}
J_{2n}(x)\cos(2n \theta)\nonumber \\
 \exp\{\lambda
\cos(\theta)\}&=& \sum_{n=-\infty}^{\infty}I_n(\lambda)\exp(i n
\theta).
\end{eqnarray}

\section{Two-time Fourier transform}\label{app:fourier}
The two-time Fourier transform has been defined according to the
following definitions:
\begin{eqnarray}
g(t_1,t_2)&=&\int \frac{dE_1}{2\pi}\frac{dE_2}{2\pi}\times
\nonumber \\ &\times& g(E_1,E_2)\exp\{ -i E_1 t_1+i E_2 t_2
\}\nonumber \\ g(E_1,E_2)&=&\int dt_1 dt_2 g(t_1,t_2)\exp\{ i E_1
t_1-i E_2 t_2\}.
\end{eqnarray}

\section{Approximation of $J_0(x)$, $J_1(x)$, $I_0(x)$, $I_1(x)$ for $x\approx
0$.}\label{app:bessel-approx}

Throughout the paper the following approximations have been
exploited:
\begin{eqnarray}
J_0(x)&\simeq &1-\frac{x^2}{4}+\mathcal{O}(x^3)\nonumber \\
J_1(x)&\simeq &\frac{x}{2}+\mathcal{O}(x^3)\nonumber \\
I_0(x)&\simeq &1+\frac{x^2}{4}+\mathcal{O}(x^3)\nonumber \\
I_1(x)&\simeq &\frac{x}{2}+\mathcal{O}(x^3).
\end{eqnarray}


\end{document}